\documentclass[amsmath,amssymb]{revtex4}
\usepackage{txfonts}%
\usepackage{graphicx}
\usepackage{epsfig}
\usepackage{graphics,color}
\usepackage{amssymb}
\usepackage{amsbsy}

 \newcommand{\be}{\begin{equation}}
 \newcommand{\ee}{\end{equation}}
 \newcommand{\bse}{\begin{subequations}}
 \newcommand{\ese}{\end{subequations}}
 \newcommand{\bea}{\begin{eqnarray}}
 \newcommand{\eea}{\end{eqnarray}}

\newcommand{\bean}{\begin{eqnarray*}}
\newcommand{\eean}{\end{eqnarray*}}

\begin{document}
\title{Formulation of transverse mass distributions in Au-Au collisions at $\sqrt{\mathrm{\it s_{NN}}}=$ 200 GeV/nucleon}
\vskip1.0cm
\author{Bao-Chun Li\footnote{libc2010@163.com, s6109@sxu.edu.cn}, Ya-Zhou
Wang  and Fu-Hu Liu} \affiliation{Department of Physics, Shanxi
University, Taiyuan, Shanxi 030006, China}

\vskip2.0cm

\begin{abstract}

The transverse mass spectra of light mesons produced in Au-Au
collisions at $\sqrt{\mathrm{\it s_{NN}}}=$ 200 GeV/nucleon are
analyzed in Tsallis statistics. In high energy collisions, it has
been found that the spectra follow a generalized scaling law. We
applied Tsallis statistics to the description of different particles
using the scaling properties. The calculated results are in
agreement with experimental data of PHENIX Collaboration. And, the
temperature of emission sources is extracted consistently.
\end{abstract}

\maketitle

{\section{Introduction}}

Multiparticle production is an important experimental phenomenon at
Relativistic Heavy Ion Collider (RHIC) in Brookhaven National
Laboratory (BNL). In Au-Au collisions, identified particle yields
per unity of rapidity integrated over transverse momentum $p_T$
ranges have provided information about temperature $T$ and chemical
potential $\mu$ at the chemical freeze-out by using a statistical
investigation~\cite{Adams:2003xp}. It brings valuable insight into
 properties of quark-gluon plasma (QGP) created in the collisions. A
much broader and deeper study of QGP will be done at
 Large Hadron Collider (LHC) at the European Organization for Nuclear
Research (CERN) and the Facility for Antiproton and Ion Research
(FAIR) at the Gesellschaft f\"ur Schwerionenforschung mbH (GSI).

 In order to estimate hadronic decay backgrounds in
 photon, single lepton and dilepton spectra which are
penetrating probes of QGP, $m_T$ spectra of identified mesons have
been studied in detail~\cite{Adare:2010fe, Adler:2006bv,
Adare:2007dg, Adare:2006kf, Adare:2009js}, where
$m_T=\sqrt{m_0^2+p_T^2}$ is transverse mass of a particle with rest
mass $m_0$ at a given $p_T$. In Ref.~\cite{Albrecht:1995ug},  $m_T$
spectral shapes of pions and $\eta$ mesons in S-S and S-Au
collisions are identical. Such behaviors are caused by $m_T$ scaling
properties, which help us to predict new $m_T$ spectra and
understand the mechanism of meson production. Statistical analysis
of $m_T$ spectra is extremely useful to extract information of
particle production process and interaction in hadronic and QGP
phases.  In the CGS (Color Glass Condensate) description, the total
hadron multiplicity follows a scaling behavior motivated by the
gluon saturation.

Different phenomenological models of initial coherent multiple
interactions and particle transport have been introduced to describe
the production of  final-state particles~\cite{t1,t2} in  Au-Au
collisions. With Tsallis statistics' development and success in
dealing with non-equilibrated complex systems in condensed matter
research~\cite{Tsallis}, it has been utilized to understand the
particle production in high-energy physics~\cite{Tang:2008ud,
Shao:2009mu, Wong:2012zr}. In our previous work~\cite{liu3}, the
temperature information of emission sources was understood
indirectly  by a excitation degree, which varies with location in a
cylinder.  We have obtained emission source location dependence of
the exciting degree specifically. From central axis to side-surface
of the cylinder, the excitation degree of the emission source
decreases linearly with the direction of radius. In this work, we
parametrize experimentally measured $m_T$ spectra of pions in
Tsallis statistics. Using the $m_T$ scaling properties in the
spectrum calculation, we  reproduce $m_T$ spectra of other light
mesons and obtain the temperature of emission sources directly.

{\section{The formulation and comparison with PHENIX results}}

At the initial stage of nucleus-nucleus collisions, plenty of
primary nucleon-nucleon collisions happen.  The primary
nucleon-nucleon collision can be regarded as an emission source (a
compound hadron fireball) at intermediate energy or a few sources
(wounded partons and woundless partons) at high energy. The
participant nucleons in primary collisions have probabilities to
take part in cascade collisions with latter nucleons. Meanwhile, the
particles produced in primary or cascade nucleon-nucleon collisions
have probabilities to take part in secondary collisions with latter
nucleons and other particles. Each cascade (or secondary) collision
is also regarded as an emission source or a few sources. Many
emission sources of final-state particles are expected to be formed
in the collisions.

 According to Tsallis statistics~\cite{Tsallis}, the total
number of  the mesons is given by \bea N=gV \int
\frac{d^3P}{(2\pi)^3}\left[1+(q-1)\frac{E-\mu}{T}\right
]^{-q/(q-1)},\label{b} \eea where $p$, $E$, $T$, $\mu$, $V$ and $g$
are the momentum, the energy, the temperature, the chemical
potential,  the volume and the degeneracy factor, respectively,  a
parameter $q$ is used to characterize the degree of nonequilibrium.
The corresponding momentum distribution is
 \bea E\frac{d^3N}{d^3P}=
\frac{gVE}{(2\pi)^3}\left[1+(q-1)\frac{E-\mu}{T}\right
]^{-1/(q-1)}.\label{b}
 \eea
We have the transverse mass $m_T$ distribution,

 \bea \frac{d^2N}{m_Tdm_Tdy}{\biggl |}_{y=0}=
\frac{gVm_T\cosh{y}}{(2\pi)^2}\left[1+(q-1)\frac{m_T\cosh{y}}{T}\right
]^{-1/(q-1)}.\label{b}
 \eea
 At midrapidity $y=0$, for zero chemical potential, the transverse
mass spectrum in terms of $y$ and $m_T$ is
 \bea \frac{d^2N}{m_Tdm_Tdy}{\biggl |}_{y=0}=
\frac{gVm_T}{(2\pi)^2}\left[1+(q-1)\frac{m_T}{T}\right
]^{-1/(q-1)},\label{b}
 \eea
which is only a $m_T$ distribution of particles emitted in the
emission source at midrapidity $y=0$.

Considering a width of the corresponding rapidity distribution of
final-state particles, the $m_T$ spectrum is rewritten as \bea
\frac{dN}{m_Tdm_T} =C\int_{-Y}^Y\cosh{y}dym_T
\left[1+(q-1)\frac{m_T\cosh{y}}{T}\right ]^{-1/(q-1)},\label{b}
 \eea
where $C=\frac{gV}{(2\pi)^2}$ is a normalization constant and
$Y$($-Y$) is the maximum (minimum) value of the observed rapidity.
Generally speaking, the temperature $T$ and $q$ can be fixed for
different event centralities (or impact parameters) by fitting
 the experimental data of pions. The temperature $T$ of
 emission sources is calculated naturally and consistently  in the current formulation.

Fig. 1 shows $m_T$ distributions of charged and neutral pions in
Au-Au collisions at $\sqrt{\mathrm{\it s_{NN}}}=$ 200 GeV/nucleon.
The symbols represent experimental data of PHENIX
Collaboration~\cite{Adare:2008qa, Adler:2003cb} and the curves are
fitting results by using Eq.(5). By fitting the experimental data,
 values of $T$ and $q$ are given in Table I with $\chi^2$/dof (degree
of freedom). Fig. 2, Fig. 3 and Fig. 4 present invariant yields of
$K^\pm $, $J/\psi$, $\phi$, $\omega$ and $\eta$ as a function of
$m_T-m_0$. The symbols represent experimental data of
 PHENIX Collaboration~\cite{Adler:2003cb, Adare:2006ns,
Adare:2010pt, Adare:2011ht, Adler:2006hu, Adare:2010dc}. The curves
are calculated results by using $m_T$ scaling properties. The
corresponding $\chi^2$/dof is given in Table II. One can see that
the transverse mass scaled results of different mesons are in
agreement with the experimental data in the whole observed $m_T-m_0$
region.

The transverse momentum spectra of charged and neutral pions for
$0-100\%$, $0-20\%$, $20-60\%$  and $60-92\%$ centralities in Au-Au
collisions at $\sqrt{\mathrm{\it s_{NN}}}=200$ GeV/nucleon are shown
in Fig. 5. The different symbols are experimental
data~\cite{Adare:2008qa, Adler:2003cb} in different centrality cuts
indicated in the figure. The curves are the results obtained by
fitting the data. We fit the spectra using Tsallis distributions and
obtain the values of $T$ and $q$ which are given in Table I with
$\chi^2$/dof. It is found that the temperature $T$  increase with
the increase of the centrality. Fig. 6 and Fig. 7 show invariant
yields of $K^\pm $, $J/\psi$, $\phi$, $\omega$ and $\eta$ as a
function of $m_T-m$ in corresponding centrality cuts. The symbols
represent experimental data of PHENIX
Collaboration~\cite{Adler:2003cb, Adare:2006ns, Adare:2010pt,
Adler:2006hu, Adare:2010dc}. The curves are the results calculated
by using $m_T$ scaling properties. The values of $\chi^2$/dof  are
shown in Table II. For different centralities, the $m_T$ scaled
results are in agreement with the experimental data of different
mesons.

\begin{table}[ht]
\vspace*{-0.0cm} {\small \vspace*{-0.1cm} \caption{Values of the
parameters $T$ and  $q$ for pions in our calculations.}%
\label{table-all}%
\vspace*{-0.15cm}
\begin{center}
{\begin{tabular}{c c c c} \hline\hline
 \,\,\,\,Centrality\,\,\,\, & \,\,\,\,\,\,\,\,$T (GeV)$\,\,\,\,\,\,
&\,\,\,\, $q$ & \,\,\,\,\,\,$\chi^2$/dof\,\,\,\,\,\, \\
\hline
Minimum Bias     \,\,&$0.064\pm0.02$  \,\,\,\,& $1.094\pm0.02$  \,\,\,\,& $0.80$\,\,\\
0-20\%           \,\,&$0.078\pm0.03$  \,\,\,\,& $1.086\pm0.02$  \,\,\,\,& $0.56$\,\,\\
20-60\%          \,\,&$0.072\pm0.02$  \,\,\,\,& $1.091\pm0.03$  \,\,\,\,& $0.49$\,\,\\
60-92\%          \,\,&$0.059\pm0.03$  \,\,\,\,& $1.098\pm0.02$  \,\,\,\,& $0.38$\,\,\\
\hline\hline
\end{tabular}}
\end{center}}

\end{table}

\begin{table}[ht]
\vspace*{-0.0cm} {\small \vspace*{-0.1cm} \caption{Values of $\chi^2$/dof for Fig.2$ - $Fig.4, Fig.6 and Fig.7.}%
\label{table-all}%
\vspace*{-0.15cm}
\begin{center}
{\begin{tabular}{c c c c c} \hline\hline
 \,\,\,\,Mesons\,\,\,\, & \,\,\,\,\,\,$0-100\%$\,\,\,\,
&\,\,\,\,$0-20\%$ & \,\,\,\,\,\,\,\,\,\,$20-60\%$ ($20-40\%$ for $J/\psi$)\,\,\,\,&\,\,\,\,$60-92\%$ ($40-92\%$ for $J/\psi$)\,\,\,\,\,\\
\hline

$\,\,\,\,K^+$      \,\,&$0.42$ & \,\,\,\,$0.44$\,\,& $0.35$\,\,& $0.55$\,\,\\
$\,\,K^-$      \,\,&$0.51$ & \,\,\,\,$0.55$\,\,& $0.57$\,\,& $0.65$\,\,\\
$\,\,J/\psi$   \,\,&$0.57$ & \,\,\,\,$0.70$\,\,& $0.61$\,\,& $0.58$\,\,\\
$\phi$     \,\,&$0.59$ & \,\,\,\,$0.50$\,\,& $0.55$\,\,& $0.45$\,\,\\
$\eta$     \,\,&$0.89$ & \,\,\,\,$0.94$\,\,& $0.82$\,\,& $0.76$\,\,\\
$\omega$   \,\,&$0.70$ & \,\,\,\,$ - $\,\,& $-$\,\,& $ - $\,\,\\
\hline\hline
\end{tabular}}
\end{center}}

\end{table}

\begin{table}[ht]
\vspace*{-0.0cm} {\small \vspace*{-0.1cm} \caption{Meson-pion yield ratios.}%
\label{table-all}%
\vspace*{-0.15cm}
\begin{center}
{\begin{tabular}{c c c c c} \hline\hline
 \,\,\,\,Centrality\,\,\,\, & \,\,\,\,\,\,$K/\pi^0$\,\,\,\,
&\,\,\,\,$\eta/\pi^0$ & \,\,\,\,\,\,\,\,\,\,$\phi/\pi^0$\,\,\,\,&\,\,\,\,$(J/\psi)/\pi^0$\,\,\,\,\,\\
\hline
Minimum Bias                         \,\,&$0.531\pm0.003$ & \,\,\,\,$0.524\pm0.038$\,\,& $0.342\pm0.015$\,\,& $0.0034\pm0.0004$\,\,\\
0-20\%                               \,\,&$0.499\pm0.004$ & \,\,\,\,$0.538\pm0.020$\,\,& $0.405\pm0.012$\,\,& $0.0031\pm0.0006$\,\,\\
$20-60\%$ ($20-40\%$ for $J/\psi$)    \,\,&$0.495\pm0.003$ & \,\,\,\,$0.577\pm0.021$\,\,& $0.390\pm0.010$\,\,& $0.0042\pm0.0005$\,\,\\
$60-92\%$ ($40-92\%$ for $J/\psi$)    \,\,&$0.481\pm0.005$ & \,\,\,\,$0.545\pm0.025$\,\,& $0.304\pm0.017$\,\,& $0.0034\pm0.0005$\,\,\\
\hline\hline
\end{tabular}}
\end{center}}

\end{table}

{\section{DISCUSSIONS AND CONCLUSIONS}}

The transverse mass spectra of  mesons produced in Au-Au collisions
at $\sqrt{\mathrm{\it s_{NN}}}$ = 200 GeV/nucleon have been
investigated in the framework of Tsallis statistics. We propose a
formula for describing the distributions and fit experimental data
of pions to estimate $q$ and the temperature $T$. Using the $m_T$
scaling properties, other meson spectra are obtained and compared
with experimental data at different collision centralities. The
maximum value of $\chi^2$/dof is 0.96, and the minimum value of
$\chi^2$/dof is 0.38. It is demonstrated that our results agree well
with the available experimental data. The normalization parameters
in the fits give the values of meson-pion yield ratio in Table III.
The ratios are helpful to understand the contribution of hadronic
decay in photonic and leptonic channels.

Final-state particles produced in high-energy nuclear collisions
have attracted much attention, since attempt have been made to
understand the properties of strongly coupled QGP by studying the
possible production mechanisms ~\cite{Andronic:2009qf,
Andronic:2009qf2}. Thermal-statistical models have been successful
in describing particle yields in various systems at different
energies~\cite{liu3, liu1, liu2, Huang:2003jv}. The temperature $T$
of emission sources is very important for understanding the matter
evolution in Au-Au collisions at RHIC. In the rapidity space,
different sources of final-state particles stay at different
positions due to stronger longitudinal
flow~\cite{BraunMunzinger:1995bp, BraunMunzinger:1994xr,
Feng:2011zze} . In our previous work, we have studied the transverse
momentum spectra of strange particles produced in Cu+Cu and Au+Au
collisions at $\sqrt{\mathrm{\it s_{NN}}}$ = 62.4 and 200
GeV/nucleon  in the framework of the cylinder model, which is
developed from the fireball model.
 The temperature $T$ of emission sources was characterized indirectly by the excitation
 degree, which varies with location in the cylinder. From
central axis to side-surface of the cylinder, the excitation degree
of the emission source decreases linearly with the direction of
radius.  In the present work, we can directly extract the
temperature by using  the $m_T$ scaling law in Tsallis statistics.
The temperature $T$  increase with the increase of the centrality.
It is consistent with results obtained in our previous work. But,
the values of $T$ is given specifically in the formulation.

Summarizing up, the transverse mass distributions of mesons produced
in Au-Au at RHIC energies have been studied in Tsallis statistics,
which reproduces $m_T$ spectra of mesons by using $m_T$ scaling
properties. The $m_T$ scaled spectra for each meson are compared
with experimental data of  PHENIX Collaboration. The formulation is
successful in the description of meson production. At the same time,
it can offer information about $m_T$-scaling properties and
the temperature of emission sources in the collisions.\\

 {\bf Acknowledgments.}  This work is supported by the
National Natural Science Foundation of China under Grant no.
11247250, no. 11005071 and no. 10975095, the National Fundamental
Fund of Personnel Training under Grant no. J1103210, the Shanxi
Provincial Natural Science Foundation under Grant no. 2013021006 and
no. 2011011001, the Open Research Subject of the Chinese Academy of
Sciences Large-Scale Scientific Facility under Grant no. 2060205,
and the Shanxi Scholarship Council of China.

\newpage

\newpage

\begin{figure}[th]
\includegraphics[width=0.7\textwidth] {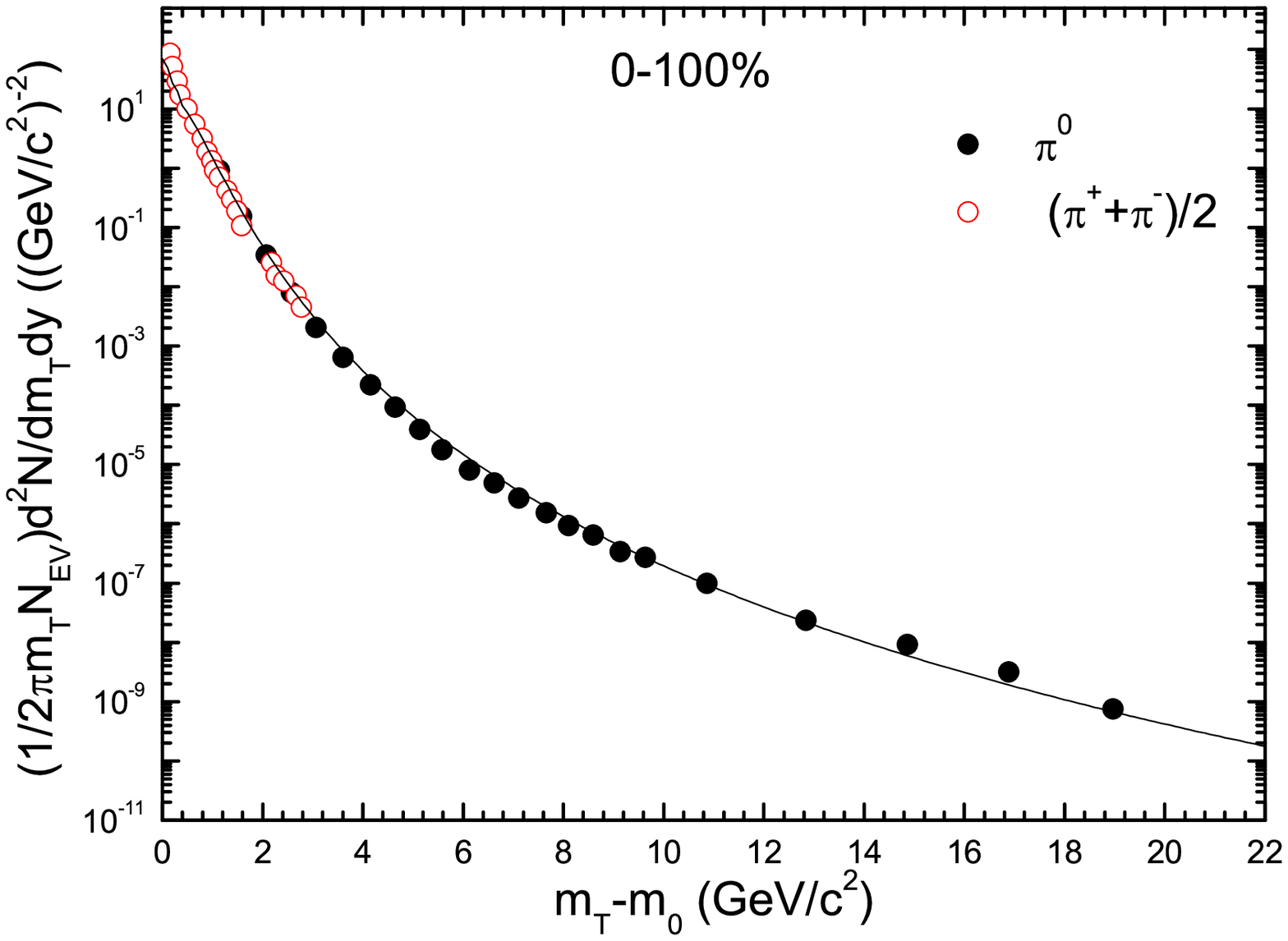}

\caption{Pion transverse mass spectra in $\sqrt{\mathrm{\it
s_{NN}}}$ = 200 GeV/nucleon Au-Au collisions. Experimental data are
taken from PHENIX Collaboration~\cite{Adare:2008qa, Adler:2003cb},
and are shown with the scattered symbols. Our calculated results are
shown with the curves.}
 \label{S1L}
\end{figure}

\begin{figure}[h]
\includegraphics[width=0.7\textwidth] {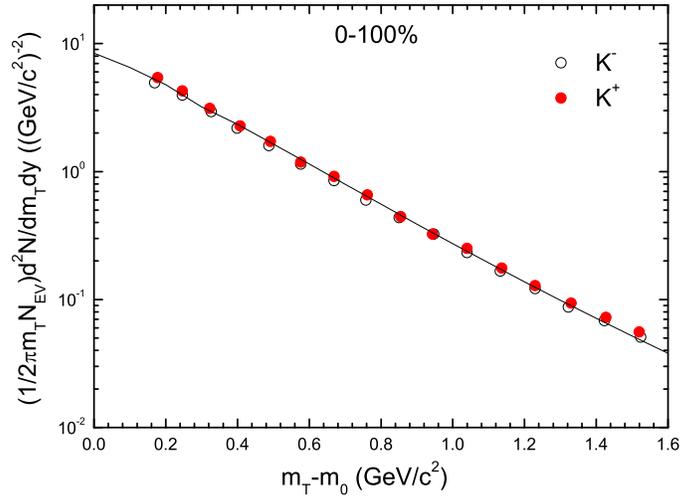}

\caption{$K^-$ and $K^+$ transverse mass spectra in
$\sqrt{\mathrm{\it s_{NN}}}$ = 200 GeV/nucleon Au-Au collisions.
Experimental data are taken from  PHENIX
Collaboration~\cite{Adler:2003cb}, and are shown with the scattered
symbols. Transverse mass scaled results are shown with the curves.}
 \label{S2L}
\end{figure}

\begin{figure}[h]
\includegraphics[width=0.45\textwidth] {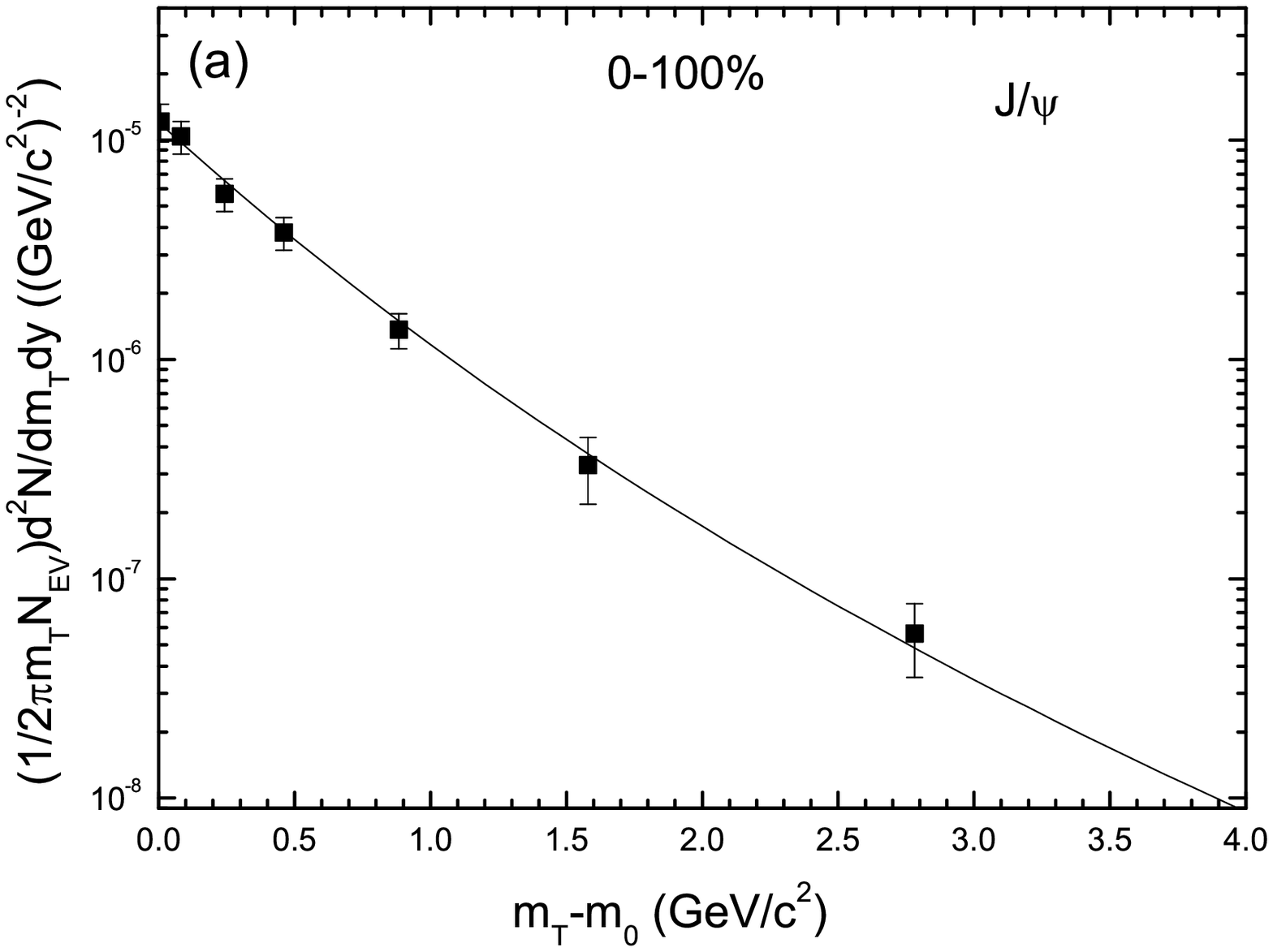}
\includegraphics[width=0.45\textwidth] {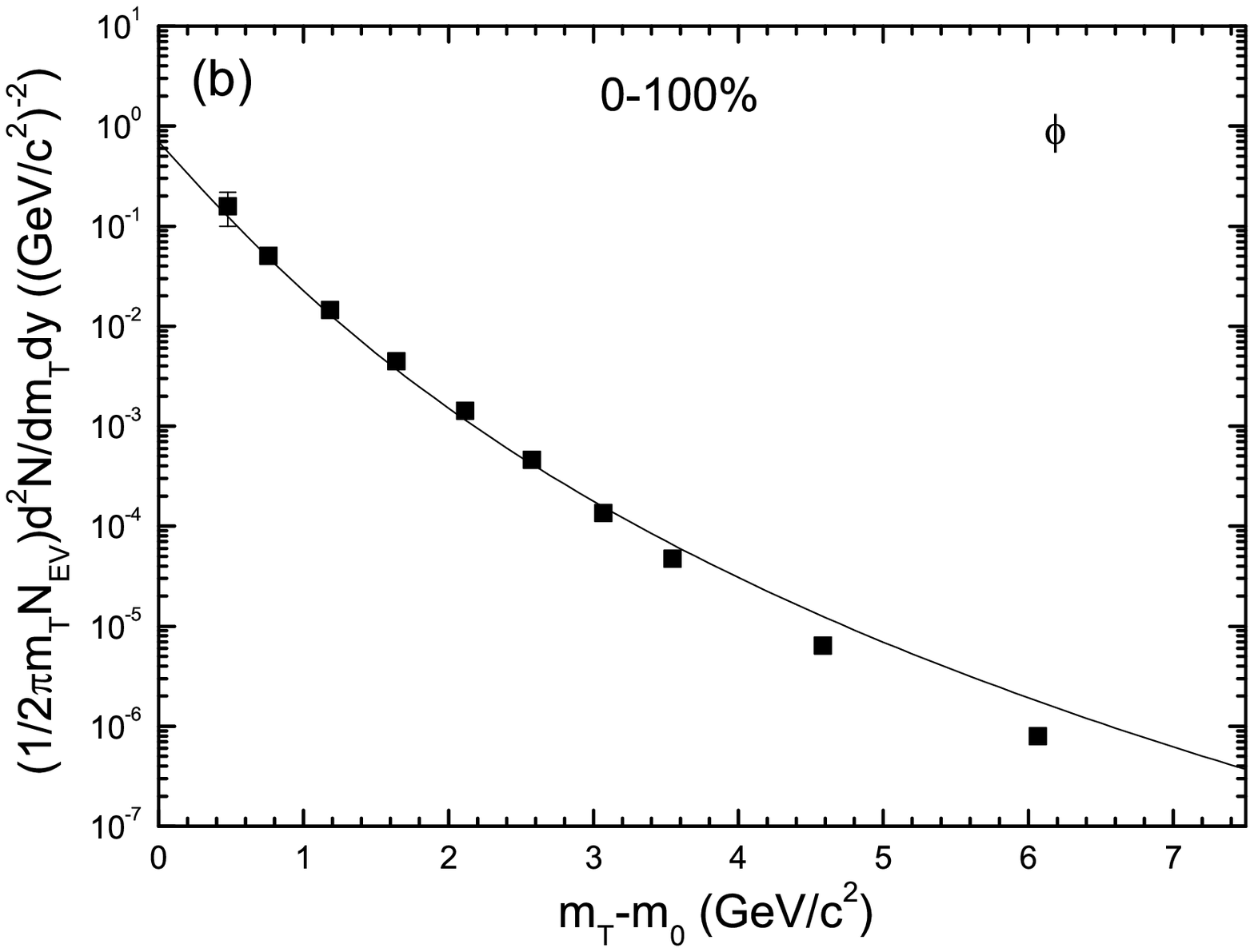}

\caption{$J/\psi$ and $\phi$ transverse mass spectra in
$\sqrt{\mathrm{\it s_{NN}}}$ = 200 GeV/nucleon Au-Au collisions. The
symbols represent  experimental data from the PHENIX
Collaboration~\cite{Adare:2006ns, Adare:2010pt} in different $P_T$
ranges. The curves are transverse mass scaled results.}
 \label{S1L}
\end{figure}

\begin{figure}[th]
\includegraphics[width=0.45\textwidth] {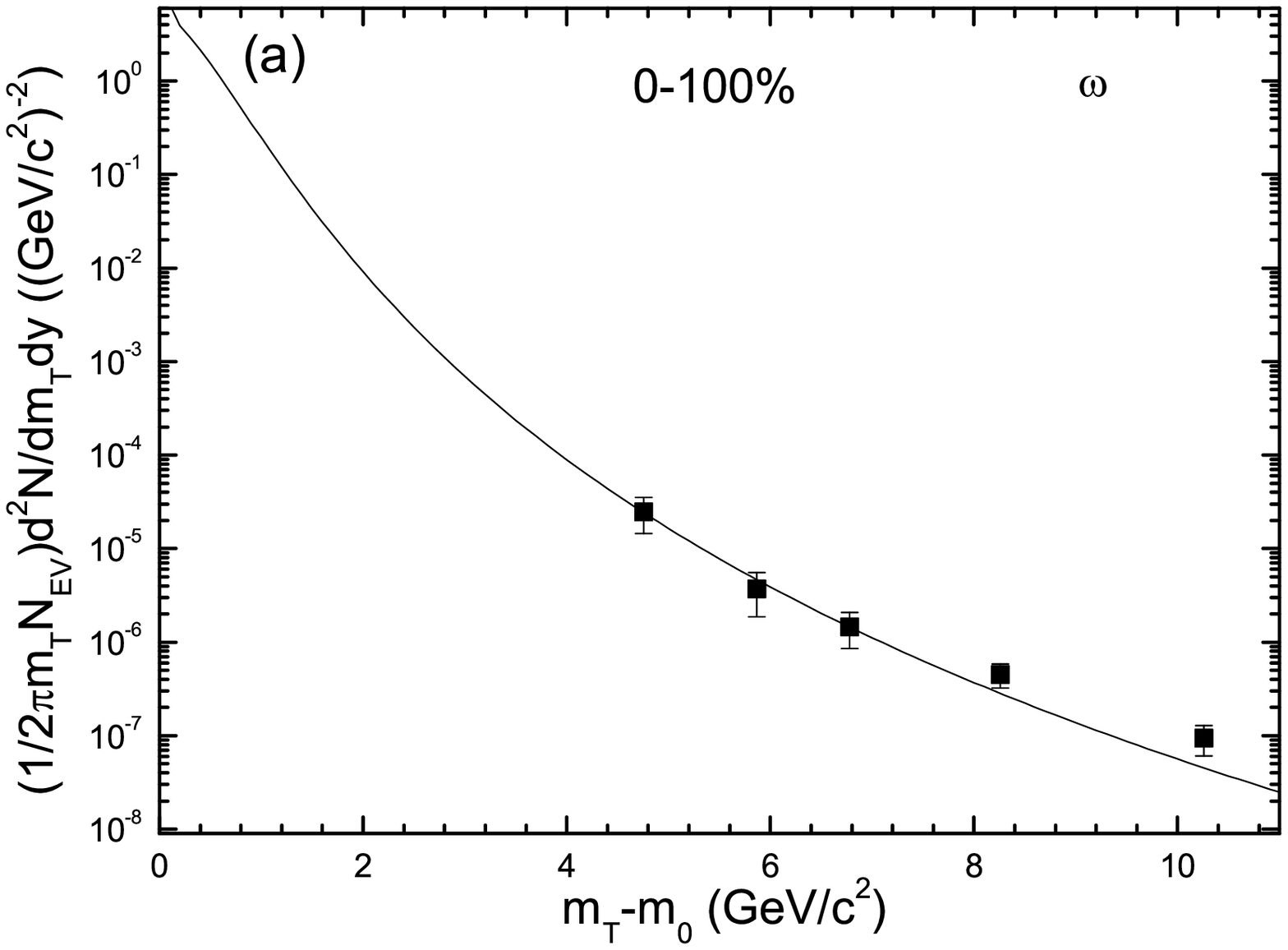}
\includegraphics[width=0.45\textwidth] {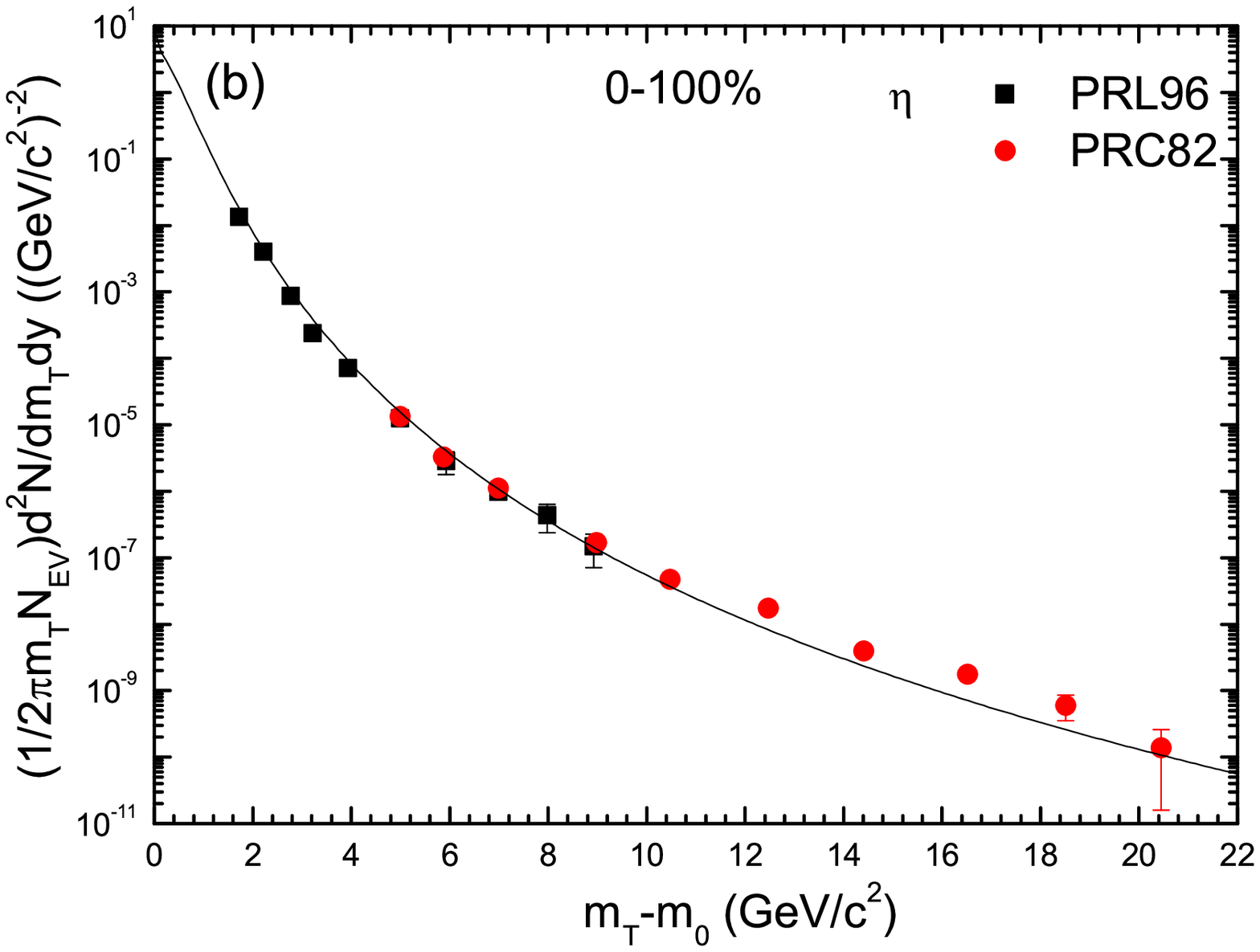}
\caption{$\omega$ and $\eta$ transverse mass spectra in
$\sqrt{\mathrm{\it s_{NN}}}$ = 200 GeV/nucleon Au-Au collisions. The
symbols represent the experimental data from the STAR
Collaboration~\cite{Adare:2011ht, Adler:2006hu, Adare:2010dc}. The
curves are transverse mass scaled results.}
 \label{S1L}
\end{figure}

\begin{figure}[h]
\includegraphics[width=0.7\textwidth] {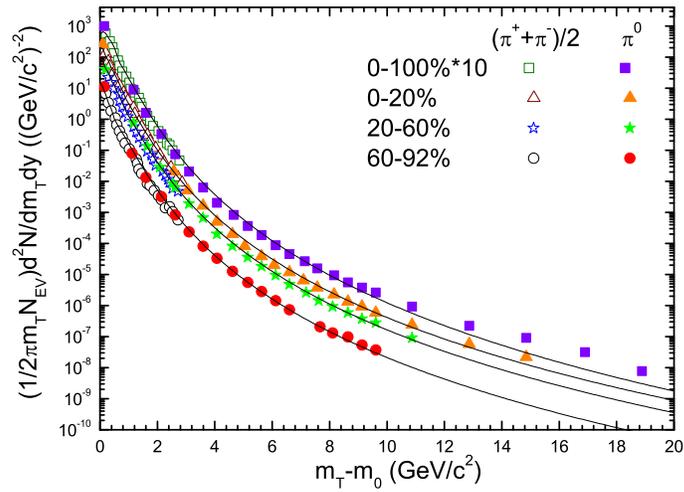}

\caption{Pion transverse mass spectra for different centrality bins
in $\sqrt{\mathrm{\it s_{NN}}}$ = 200 GeV/nucleon Au-Au collisions.
The symbols represent the experimental data from PHENIX
Collaboration~\cite{Adare:2008qa, Adler:2003cb}. Our calculated
results are shown with the curves.}
 \label{S1L}
\end{figure}

\begin{figure}[h]
\includegraphics[width=0.45\textwidth] {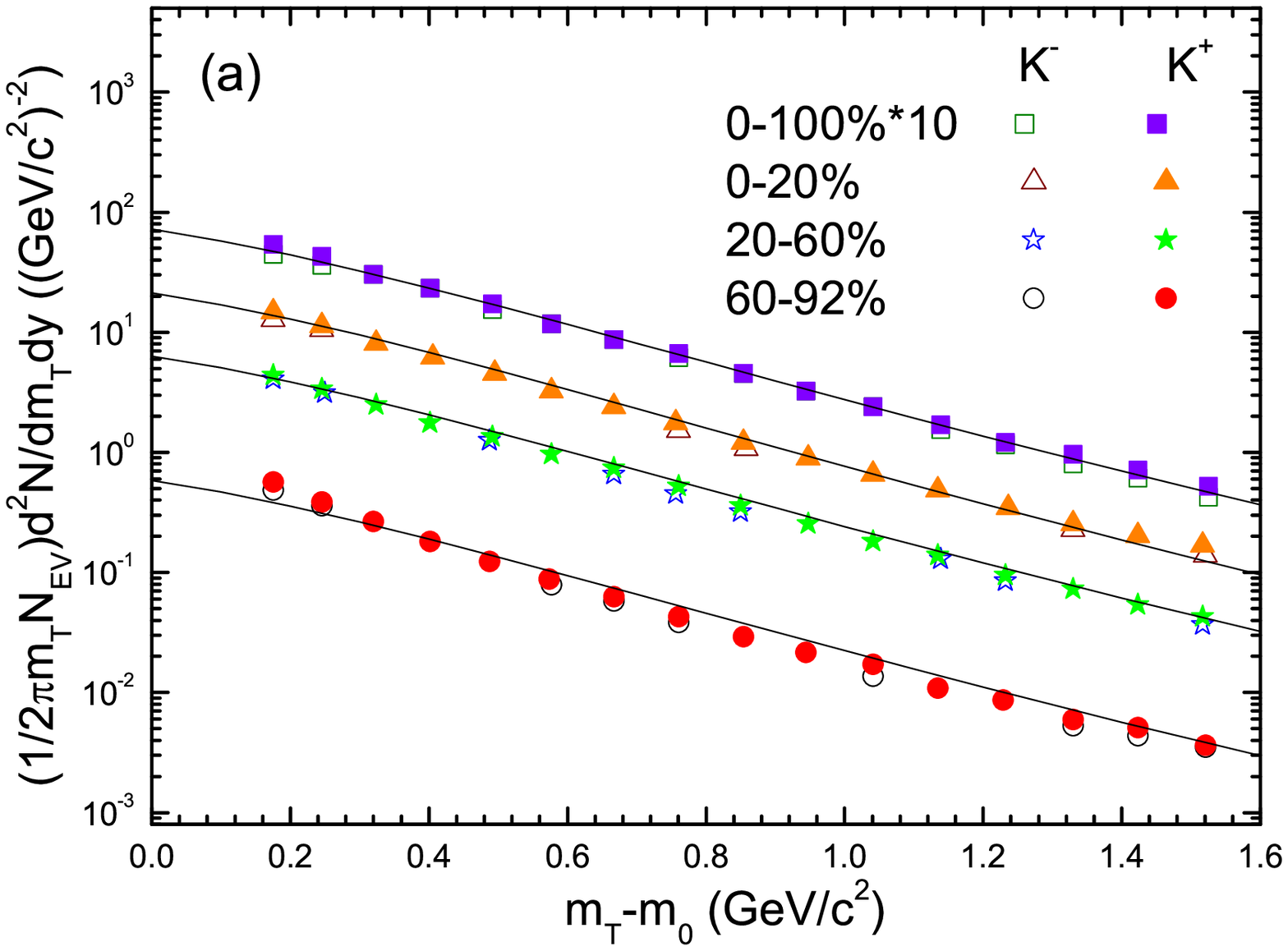}
\includegraphics[width=0.45\textwidth] {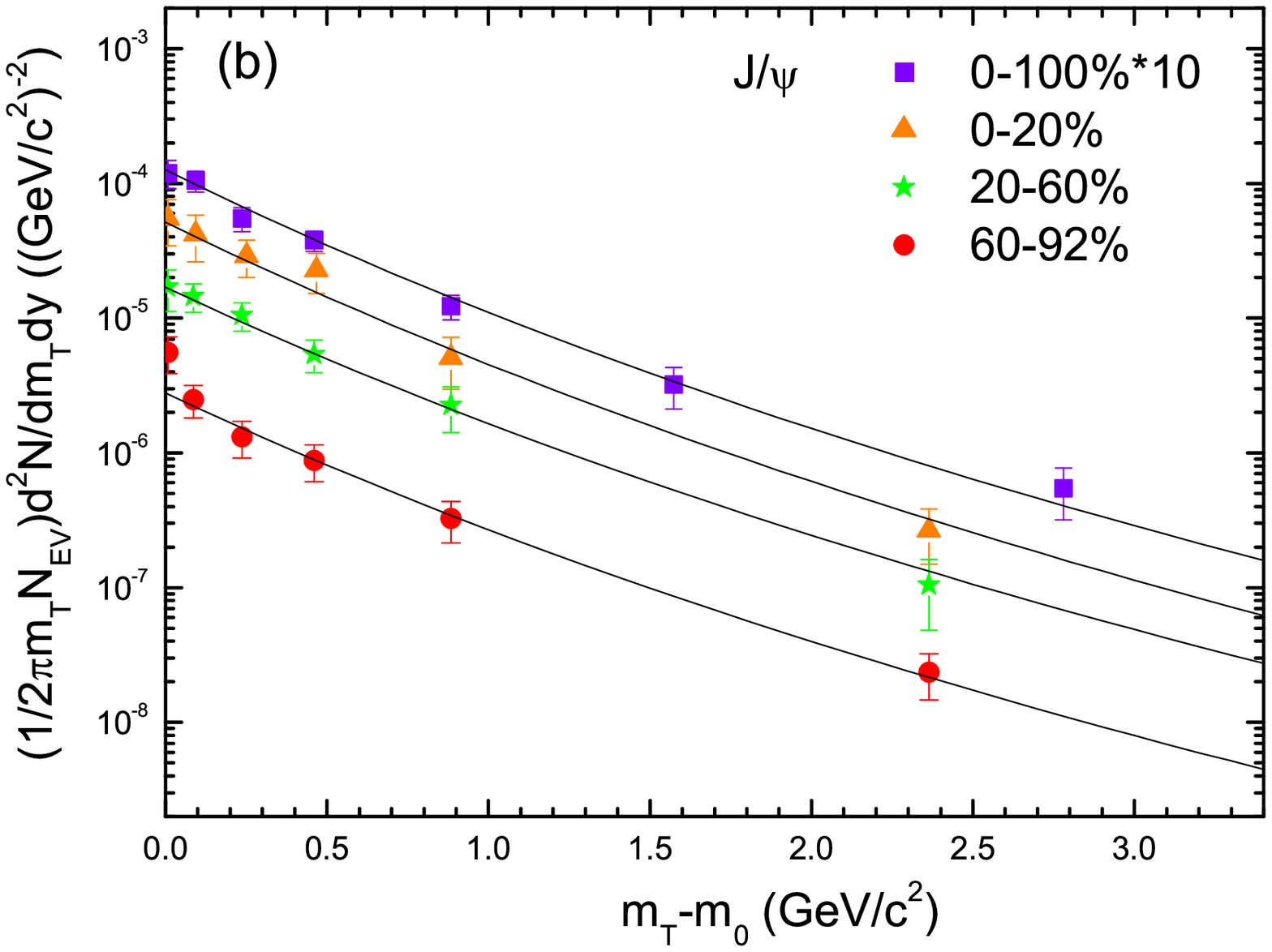}

\caption{$K^-$, $K^+$ and $J/\psi$ transverse mass spectra for
different centrality bins in $\sqrt{\mathrm{\it s_{NN}}}$ = 200
GeV/nucleon Au-Au collisions. The symbols represent the experimental
data from PHENIX Collaboration~\cite{Adler:2003cb, Adare:2006ns}.
The curves are transverse mass scaled results.}
 \label{S1L}
\end{figure}

\vspace{4cm}
\begin{figure}[htb]
\includegraphics[width=0.45\textwidth] {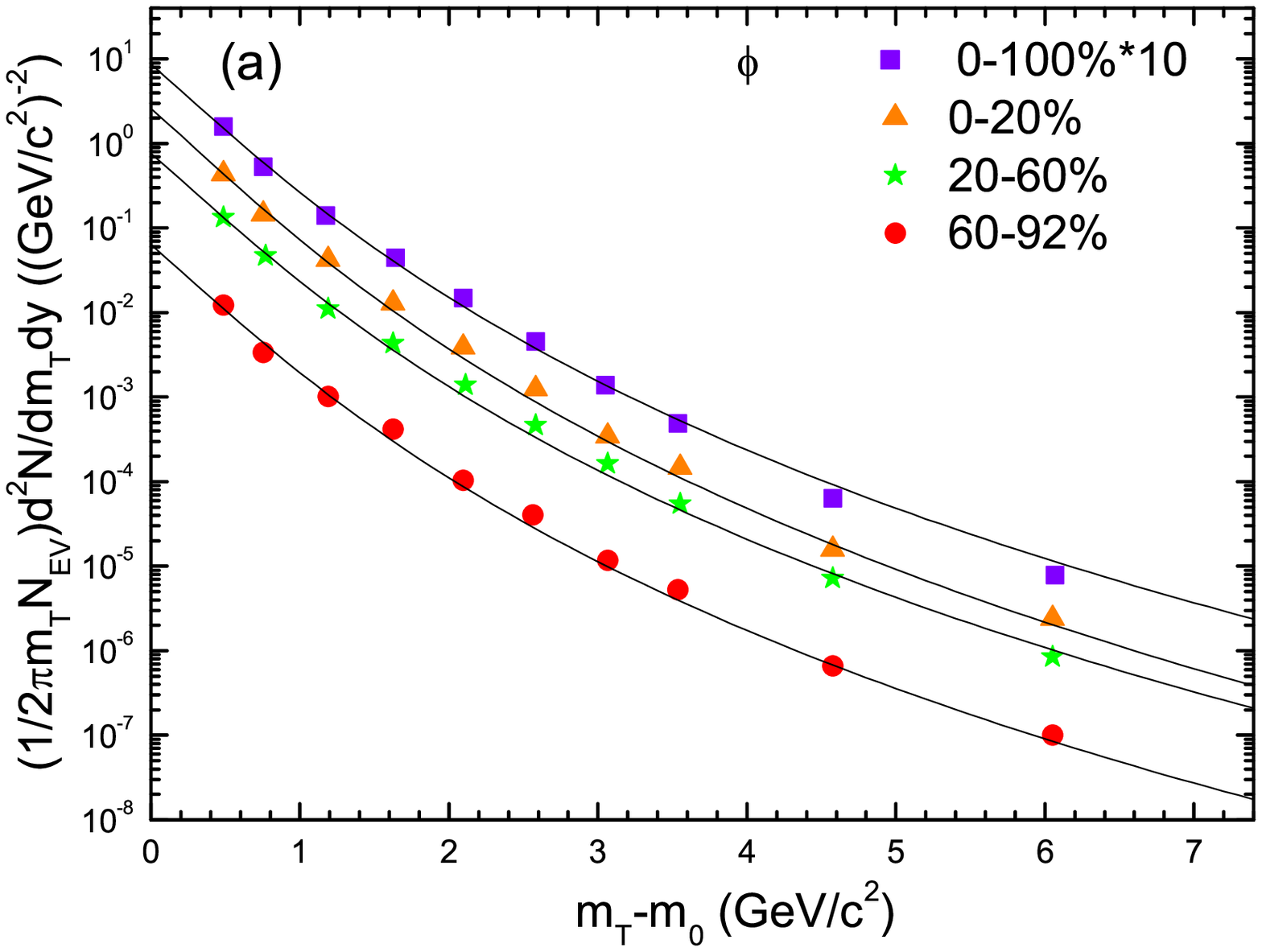}
\includegraphics[width=0.45\textwidth] {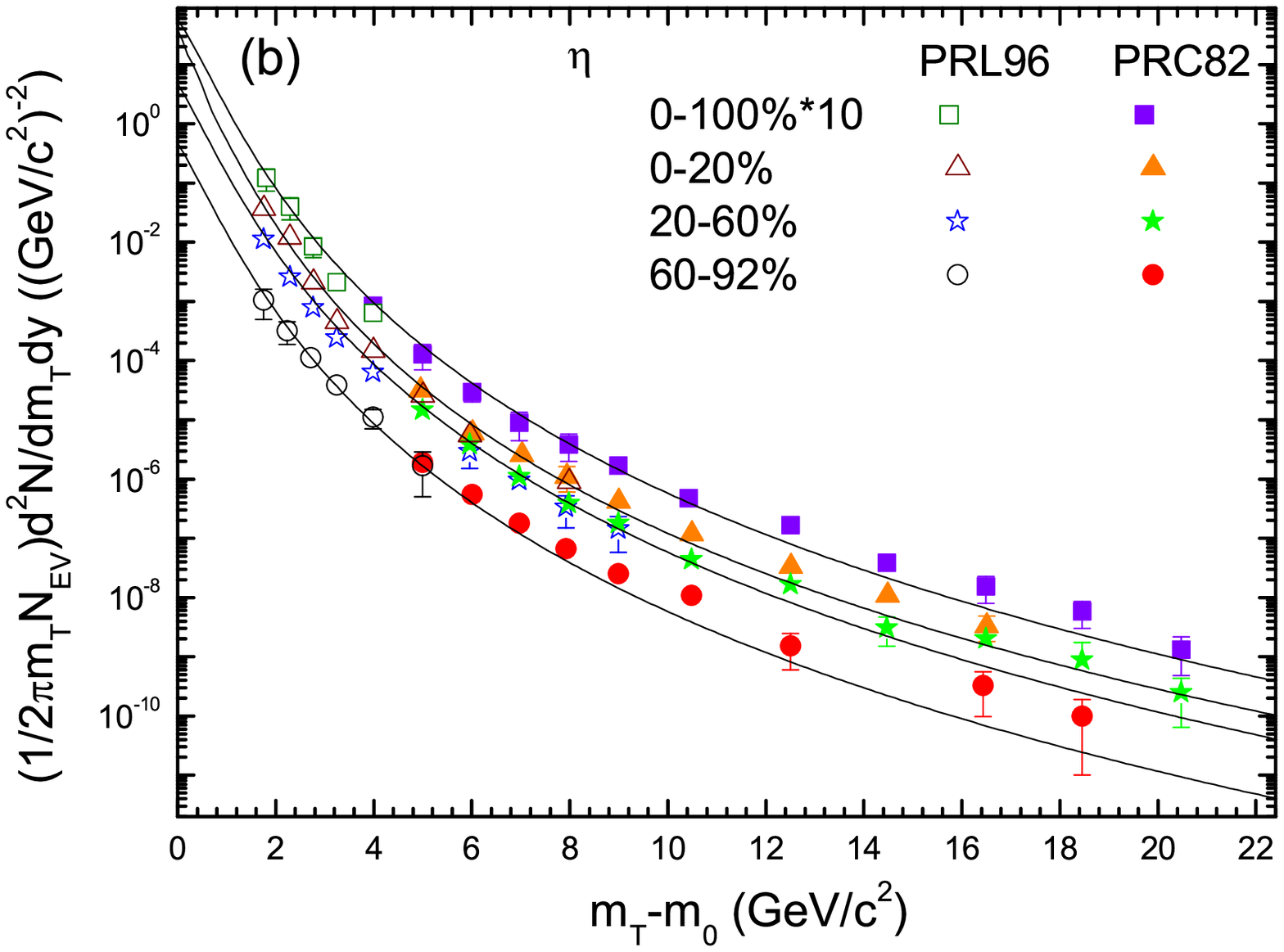}

\caption{$\phi$ and $\eta$ transverse mass spectra for different
centrality bins in $\sqrt{\mathrm{\it s_{NN}}}$ = 200 GeV/nucleon
Au-Au collisions. The symbols represent the experimental data from
PHENIX Collaboration~\cite{Adare:2010pt, Adler:2006hu,
Adare:2010dc}. Transverse mass scaled results are shown with the
curves.}
 \label{S1L}
\end{figure}

\end{document}